\documentclass[aps,prl,twocolumn,superscriptaddress]{revtex4}
\pdfoutput=1
\usepackage{graphicx}
\usepackage{bm}
\usepackage{amsmath}
\usepackage{color}

\begin{document}

\preprint{tsk-etr-v10}

\title{Giant nonlinearity of carbon nanotubes in a photonic metamaterial}

\author{Andrey E. Nikolaenko}
\affiliation{Optoelectronics Research Centre, University of Southampton, Southampton SO17 1BJ, United Kingdom}

\author{Francesco De Angelis}
\affiliation{Italian Institute of Technology, 16163 Genova and the
University of Magna Graecia, 88100 Catanzaro, Italy}

\author{Stuart A. Boden}
\affiliation{School of Electronics and Computer Science, University of Southampton, Southampton SO17 1BJ, United Kingdom}

\author{Nikitas Papasimakis}
\affiliation{Optoelectronics Research Centre, University of Southampton, Southampton SO17 1BJ, United Kingdom}

\author{Peter Ashburn}
\affiliation{School of Electronics and Computer Science, University of Southampton, Southampton SO17 1BJ, United Kingdom}

\author{Enzo Di Fabrizio}
\affiliation{Italian Institute of Technology, 16163 Genova and the
University of Magna Graecia, 88100 Catanzaro, Italy}

\author{Nikolay I. Zheludev}
\affiliation{Optoelectronics Research Centre, University of Southampton, Southampton SO17 1BJ, United Kingdom}

\date{\today}

\begin{abstract}
Metamaterials, artificial media structured on the
subwavelength scale offer a rich paradigm for developing unique
photonic functionalities ranging from negative index of refraction
and directionally asymmetric transmission to slowing light. Here we demonstrate that a combination of carbon
nanotubes with a photonic metamaterial offers a new paradigm for the development of nonlinear media with exceptionally strong ultrafast nonlinear response invaluable in photonic applications. It is underpinned by strong coupling between
weakly radiating Fano-type resonant plasmonic modes and the excitonic
response of single-walled semiconductor carbon nanotubes. Using a "combinatorial" approach to material discovery we show that the optical response of such a composite system can be tailored and optimized by metamaterial design.
\end{abstract}

\maketitle

Carbon nanotubes (CNTs) are nearly ideal one-dimensional systems,
with diameter of only a few nanometers and length on the micron scale. Single walled CNTs rolled from a graphene sheet to create spiral arrangements of atoms along the tube are of particular
interest to photonics. Such nanotubes are direct gap
semiconductors with absorbtion spectra dominated by exciton
lines~\cite{Excitons CNT}. Their possible technological uses
include nanometre-scale light sources, photodetectors and
photovoltaic devices. CNTs also possess unique nonlinear optical properties~\cite{Nature CNT review} as they exhibit high
third-order susceptibility with sub-picosecond recovery time
\cite{CNT fast nonlinearity, CNT fast nonlinearity 2} lending to
applications in ultrafast lasers ~\cite{CNT mode-lock fiber,
CNT mode-lock fiber 2, CNT mode-lock solid, CNT mode-lock solid
2, CNT mode-lock solid 3}. CNTs exhibit significant advantages over other materials as nonlinear media: they offer much simpler and cheaper fabrication than conventional
semiconductor nonlinear optical components, they are robust and they can be
easily integrated into optical-fibre and waveguide environments.
\begin{figure} [h!]
\includegraphics[width=0.45\textwidth]{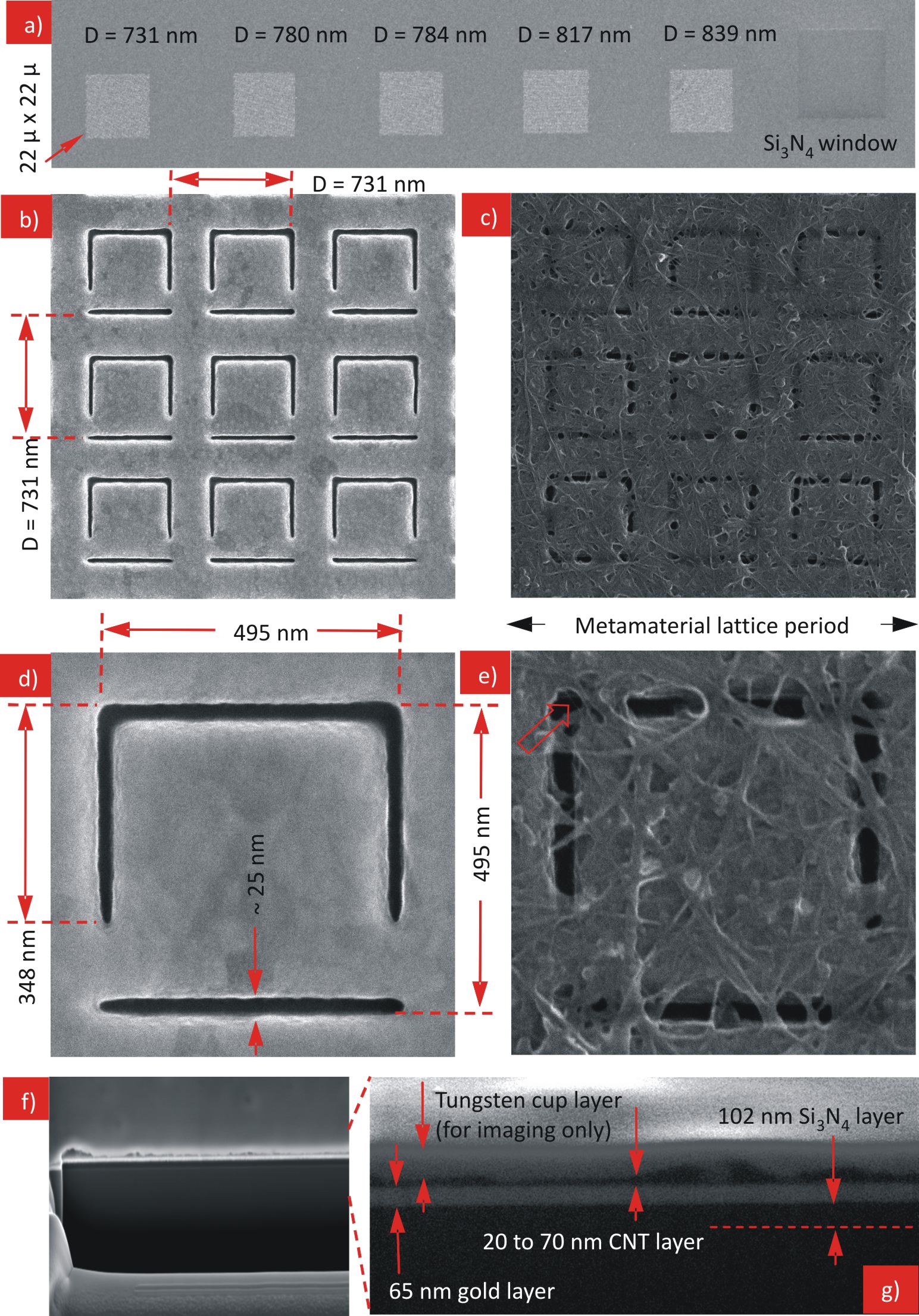}
\caption{Carbon Nanotube Metamaterial imaged under a scanning
helium ion microscope. The combinatorial sample consists of five different structurally related metamaterial designs with different unit cell sizes $D$ and an empty area annotated as "$Si{_3}N{_4}$ window" on the same substrate (a). The metamaterial structure is a two-dimensional
nanoscale array of slits in a gold film supported on a
silicon-nitride mebrane (b). Carbon nanotubes deposited on the
surface of the metal nanostructure form a layer of "nanoscale
feutre" (c). Plates (d) and (e) show unit cells of the
metamaterial before and after deposition of nanotubes. On plate
(e) note the arrow pointing at a single nanotube crossing the slit.
Plate (f) shows a milled slit manufactured to study the morphology of
the structure, which is presented in plate (g).}\label{conf}
\end{figure}

The main source of optical nonlinearity in semiconductor CNTs is
the saturation of the resonant exciton line. Combining CNTs as
nonlinearity agents with metamaterial structures provides the
opportunity to link the plasmonic resonances of metamaterials with
the excitonic resonances of nanotubes. In fact, we demonstrate here that engaging the strong local
fields in the vicinity of the metamaterial leads to an enhanced response of the nonlinear
agent. For that matter selecting an appropriate metamaterial structure is crucially important. In our
experiments we used a planar structure that belongs to the class of
metamaterials supporting dark mode plasmonic excitations~\cite{VAF
PRL}. In such metamaterials weak coupling of the excitation mode
to the free-space radiation modes creates narrow reflection, transmission and
absorption resonances with asymmetric, Fano-like dispersion. The
first example of such a metamaterial was a double-periodic array of metallic
asymmetrically split ring wire resonators that has found numerous
applications in cases where sharp spectral features are required~\cite{Eric
PRL, Vanadium, QDs, DeLaRue}. Here we used a structure
complementary to the split ring wire metamaterial: a double
periodic array of asymmetrically split ring slits in a metal film
(see Fig.~1).

The metamaterial structures were fabricated by focused ion beam
milling through a 65~nm thick gold film evaporated on a 102~nm thick
$Si{_3}N{_4}$ membrane. Gold film roughness of less than 5 nm
was obtained with low pressure ($10^{-8}$~mbar) thermal evaporation.
On a single membrane we manufactured five metamaterial arrays with
overall sizes 22~$\times$~22~$\mu$m$^2$ and different unit cell size
$D$ varying from to 731~nm to 839~nm. This allowed us to follow a
combinatorial approach for materials discovery, where a rapid search
for the optimal composition is achieved by parallel screening of a
number of different but structurally related samples~\cite{comb}.
Here the spectral position of the plasmonic resonance
$\lambda_{p}$ depends on the size of the unit cell. The available
range of metamaterial structures with different unit cell size $D$
allowed the study of the linear and nonlinear response for
varying spectral separation of the main excitonic
resonance of CNTs at $\lambda_{11}$ from the plasmonic resonance of the
metamaterial at $\lambda_{p}$, providing thus control over $\delta_{pe}=\lambda_{11}-\lambda_{p}$. For wavelengths longer
than the unit cell such periodic nanostructures do not cause
diffraction of infrared optical radiation. In the spectral range
of the CNT excitonic absorption lines they are true metamaterials as far as
their far-field electromagnetic properties are concerned and may be fully
characterized by their absorption, transmission and reflection.
Characteristic spectra of the metamaterial are presented on
Fig.~2a for an array with $D=731$~nm. Fig.~3d shows the
dependence of the peak of the metamaterial plasmon absorption line
$\lambda_p$ on the unit cell size.

\begin{figure}
\includegraphics[width=0.45\textwidth]{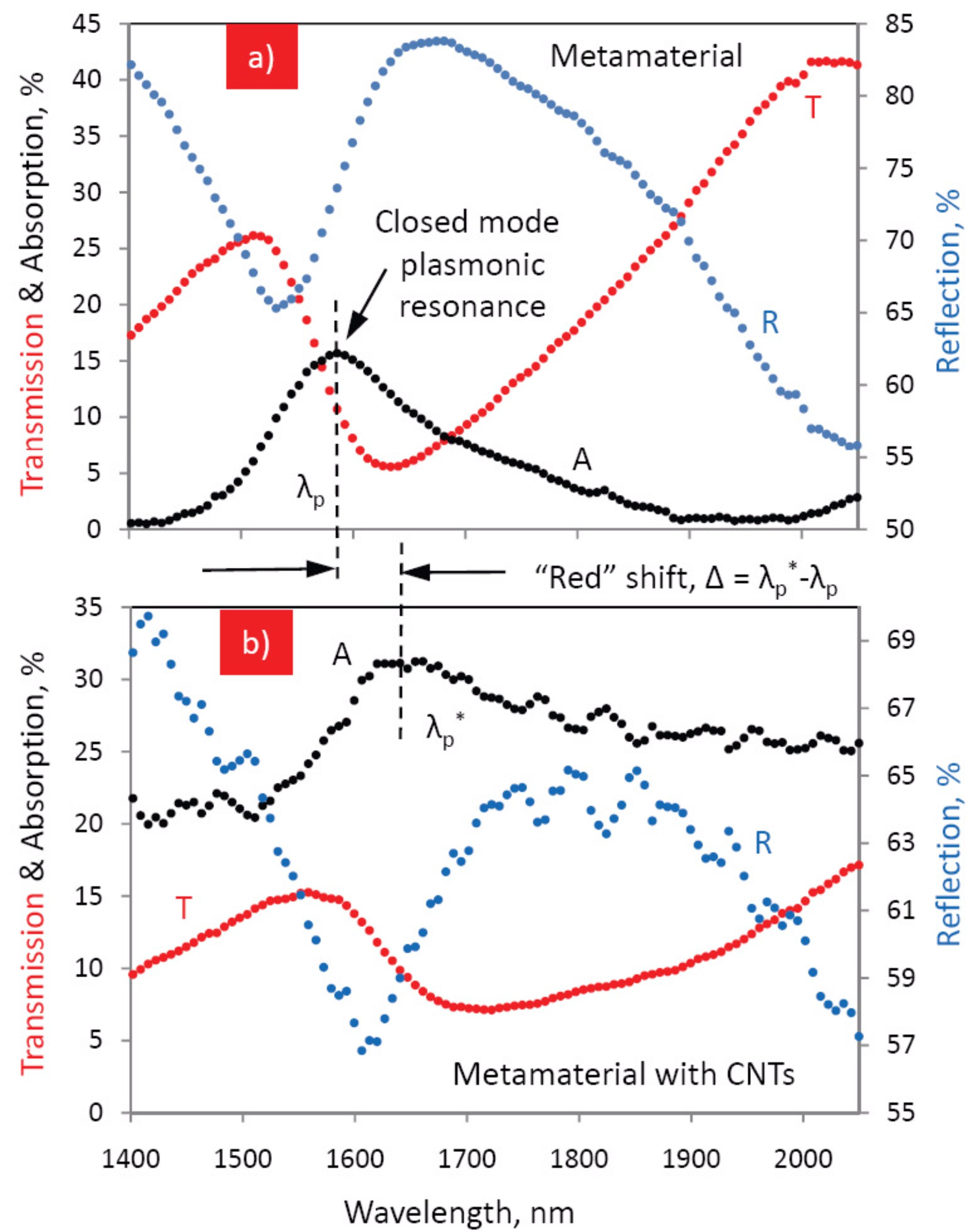}
\caption{Spectral response of the Carbon Nanotube Metamaterial: Transmission (T),
Reflection (R) and Absorption (A) for a metamaterial array with a $D=731$~nm unit cell size (a) and for the same
metamaterial fuctionalized with CNTs (b). Note the red shift of the
plasmon absorption resonance in the CNT-fuctionalized
metamaterial.}\label{ABS}
\end{figure}

The $Au@Si{_3}N{_4}$ metamaterial structures were functionalized with single walled semiconductor carbon nanotubes with a characteristic diameter of $1.4$~nm. A thin layer of nanotubes was formed on the metamaterial surface by spraying a sonificated water suspension of nanotubes and heating to $100~^0$C resulting in the rapid evaporation of the water component. When deposited on a bare $Si{_3}N{_4}$ membrane, the nanotube layer shows a characteristic absorption spectrum dominated by $\lambda_{11}$ and $\lambda_{22}$ excitonic lines, as presented in Fig.~3b. Figure 1 shows microscope images of the metamaterial before and after functionalization with CNTs. The images were taken with a scanning helium ion microscope. This novel imaging technique~\cite{He-ion Microscopy} is very well suited for imaging carbon nanotubes on metamaterials as it benefits from large depth of field, small interaction volume of ions with the medium and high contrast of the image ensuring excellent surface detail. CNTs seem to form a strongly interlinked network, a layer of "nanoscale feutre" where individual nanotubes are bunched in thicker thread-like structures. On a few occasions single nanotubes bridging the gaps of the metal nanostructure are also seen (see Fig.~1e).  We investigated the morphology of the carbon nanotube metamaterial by observing its cross-section in a trench cut through the sample by a focused ion beam (Fig.~1f). For this matter, a section of the sample was covered by a thin protective layer of tungsten. The layer of carbon nanotubes had a thickness between 20~nm and 70~nm across the sample as can be seen on Fig.~1g. It creates negligible scattering at optical frequencies as it is structured at a deep sub-wavelength scale.

\begin{figure}
\includegraphics[width=0.45\textwidth]{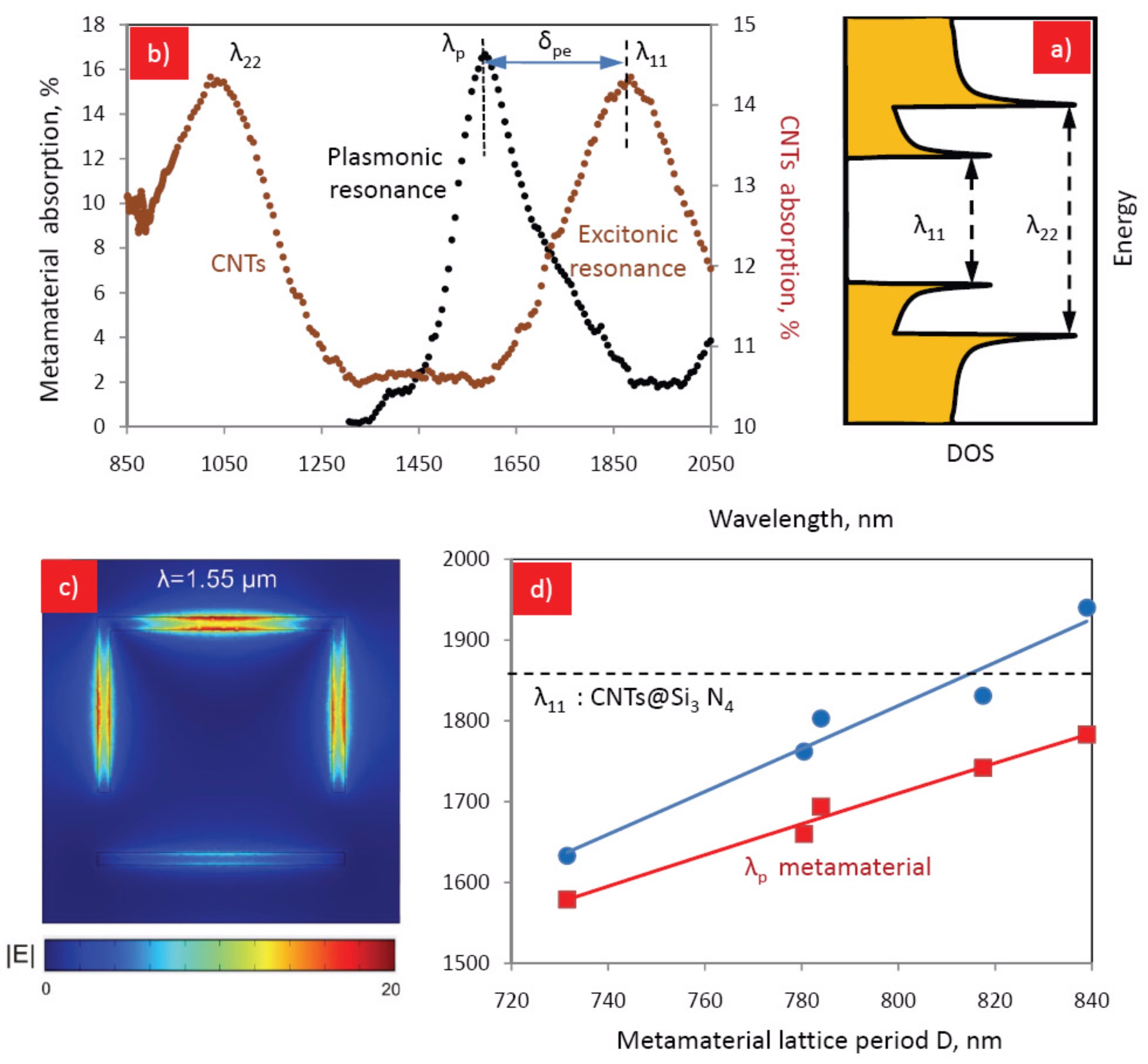}
\caption{Electronic Density of States (DOS) in a semiconductor single walled carbon nanotube (a); Plasmonic absorption resonance in a metamaterial without CNTs and excitonic resonances in a CNT film on a silicon nitride substrate (b); calculated color coded field map showing the total magnitude of the electric field of the light wave in the immediate proximity of the metamaterial plane at the plasmonic resonance $\lambda_p$ (c); and the dependence of the metamaterial absorption resonance spectral position on the unit cell size before and after functionalization with CNTs (d).}\label{SHI}
\end{figure}

We observed substantial changes in the metamaterial's optical properties resulting from the CNT functionalization. All resonance features exhibited an anticipated "red shift" $\Delta=\lambda^*_{p}-\lambda_{p}$ of the plasmon resonance resulting from the reduction of the plasmon frequency due to the presence of the highly polarizable carbon nanotubes (compare Fig.~2a and Fig.~2b). After functionalization the metamaterial's reflection decreased, whereas the spectrum of absorption accrued a background associated with the interband and exciton transition in the nanotubes. Additional losses introduced by the CNTs damped the metamaterial plasmonic resonance, hence its quality factor decreased. Hidden in the stronger spectral features of the metal nanostructure, the $\lambda_{11}$ excitonic line is not identifiable on the absorption spectrum of the functionalized metamaterial. The red-shifted positions of the trapped mode resonance in the metamaterial-CNT system are presented in Fig.~3d for different sizes of the unit cell $D$.

The nonlinear response of the metamaterial was investigated with a
broadband ultrafast super-continuum fiber source generating a
continuous train of pulses with a repetition rate of 20~MHz. The
source was equipped with a computer-controllable, tunable, 10~nm
bandwidth, spectral filter. The sample was placed at the focal
point of a dispersion-free reflective parabolic concentrator to
achieve a spot size of about 20~$\mu$m in diameter (full width at half maximum). Through
the nature of the super-continuum generation process the pulse
duration was a function of wavelength and within the investigated
spectral interval varied from a few ps to a few hundred
femtoseconds; thus it is instructive to present results of
nonlinearity measurements in terms of fluence of the light
excitation. The measurements (see Fig.~4b) were taken by increasing
fluence from about 3~$\mu J$/cm$^2$ to the level of 40~$\mu
J$/cm$^2$ corresponding to the average power level on the sample
of only about 2.4~mW.
The spectra of the nonlinear response are presented on Fig.~4. Here the
nonlinear response is normalized to the fluence level of 40~$\mu
J$/cm$^2$ across the entire spectrum. At resonance the light induced transmitted intensity variation of the carbon
nanotube metamaterial is about 10 percent.  In good agreement
with previous works \cite{CNT mode-lock solid 2, CNT mode-lock solid 3}, we detected a much weaker nonlinear
response of CNTs on the unstructured dielectric substrate (Fig.
~4a).
\begin{figure} [h!]
\includegraphics[width=0.45\textwidth]{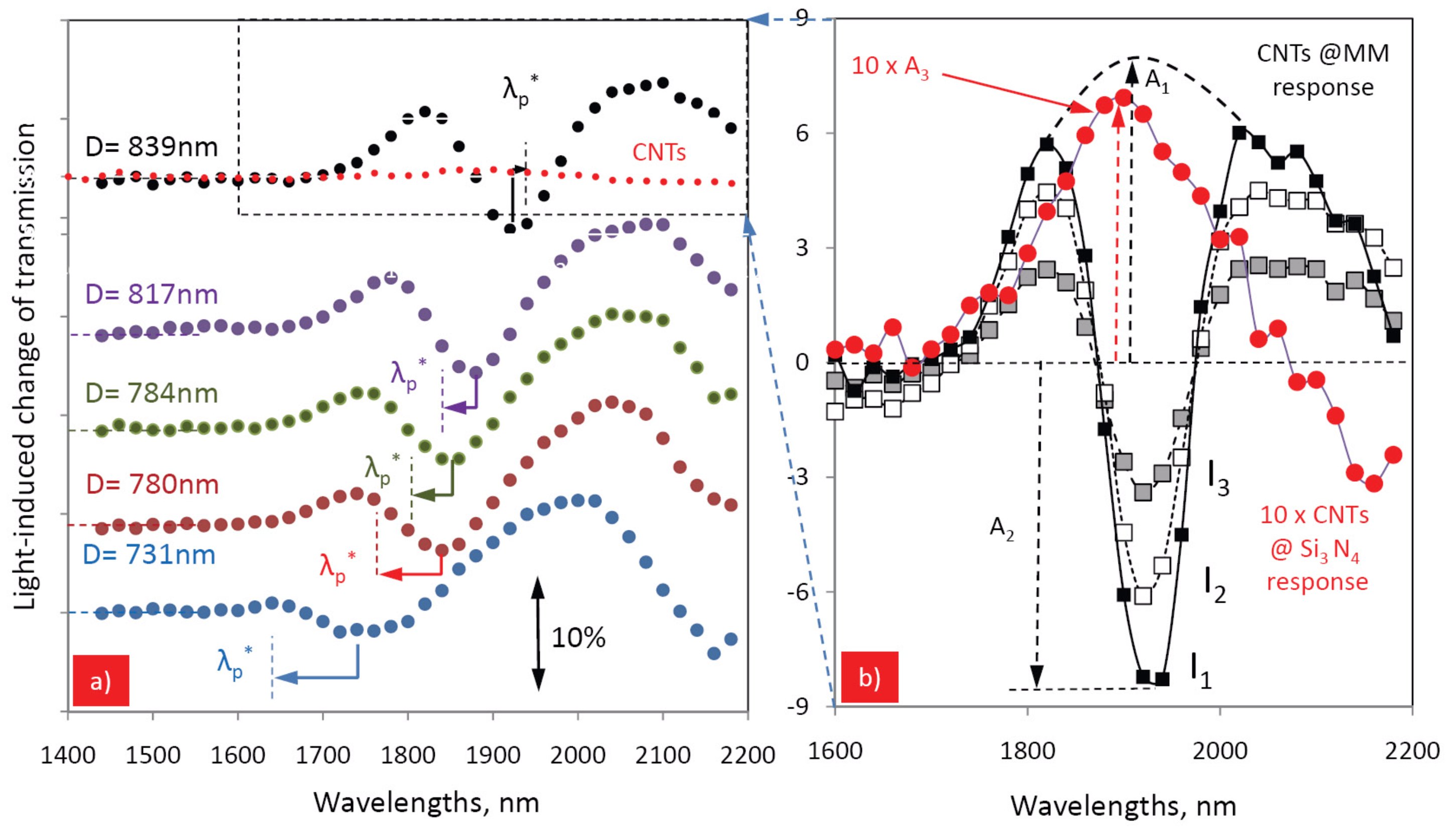}
\caption{Light-induced change of
transmission of the CNT-functionalized metamaterial for
different unit cell sizes (a). Light-induced transmission change for different
levels of intensity for a CNT-functionalized metamaterial with unit cell size $D=839$~nm (b).
The nonlinear response of CNTs on a silicon nitride membrane
(magnified by a factor of 10) is shown for comparison on panel (b) and
also on panel (a) (without magnification).}\label{NON}
\end{figure}

The nonlinear response of the metamaterial has a complex frequency
dispersion that could be decomposed on two main components
derived from the analysis of the response in structures with different
unit cell sizes. The first component that is practically
independent of the unit cell size is relevant to the bleaching of
the carbon nanotube excitonic resonance. Here increase of the
light intensity leads to an increase of transmission. On Fig.~4b
this component of the response is illustrated by a dashed
bell-shaped line with amplitude $A_1$ that is centered at the CNT's
exciton absorbtion peak at $1950$~nm and has a width of about
$310$~nm. This bleaching response is superimposed to a much
sharper "negative" peak of reduced transmission. For a
metamaterial with $D=839$~nm this peak, indicated as $A_2$, has a
width of about 120~nm. We argue that this "negative" component
is linked to the reduced damping of the plasmon mode through
exciton-plasmon coupling. Indeed under strong resonant coupling,
the lower excitonic damping results in the plasmon absorption peak becoming more intense and partially recovering the low transmission levels characteristic of the "CNT-free" metamaterial. This interpretation is
very well supported by the fact that the negative peak migrates
towards higher frequencies in structures with a smaller unit cell
size, i.e. with the reduction of the plasmon resonant wavelength.
This mechanism will be illustrated below using a classical
oscillator model.

However, as the nonlinear effect here has a transient nature, and
also involves nonlinear refraction, it takes the form of dynamic
resonance pulling, where the apparent resonance frequency of the
"negative" response is somewhere in between the excitonic line
position and the plasmonic resonance in the structure. When the
plasmon and exciton resonances nearly coincide, as in the case
presented in Fig.~4b, the "negative" effect is most pronounced. Here
the nonlinear response may be compared with that of CNTs deposited
on a bare $Si{_3}N{_4}$ membrane: one can see that the
CNT's response on the unstructured substrate (peak $A_3$) is about
12 times smaller than the overall "negative" response of the CNTs
on the metamaterial (peak $A_2$). One can also argue that the
positive response on the CNTs (peak $A_3$) is a factor of 13
smaller than that of the positive component of the CNT's response
in the metamaterial (peak $A_1$) while the overall negative
response (peak $A_1$ plus peak $A_2$) is a factor of 25 smaller
than that of the CNTs alone. From a slightly different prospective,
the enhanced non-linear response of the composite
metamaterial-carbon nanotube system is due to the resonant
increase of the plasmon fields in the vicinity of the slits in the
metal film through which light penetrates the metamaterial. This
is illustrated in Figs. 3c, where we present the results of full
three-dimensional Maxwell calculations of the electric field
magnitude just above the metamaterial surface. At the plasmon
resonance $\lambda_p$ the field just above the split-ring slits is up to 20 times
stronger than the electric field of the incident wave, ensuring a
strongly intensity-dependent response.

\begin{figure} [h!]
\includegraphics[width=0.45\textwidth]{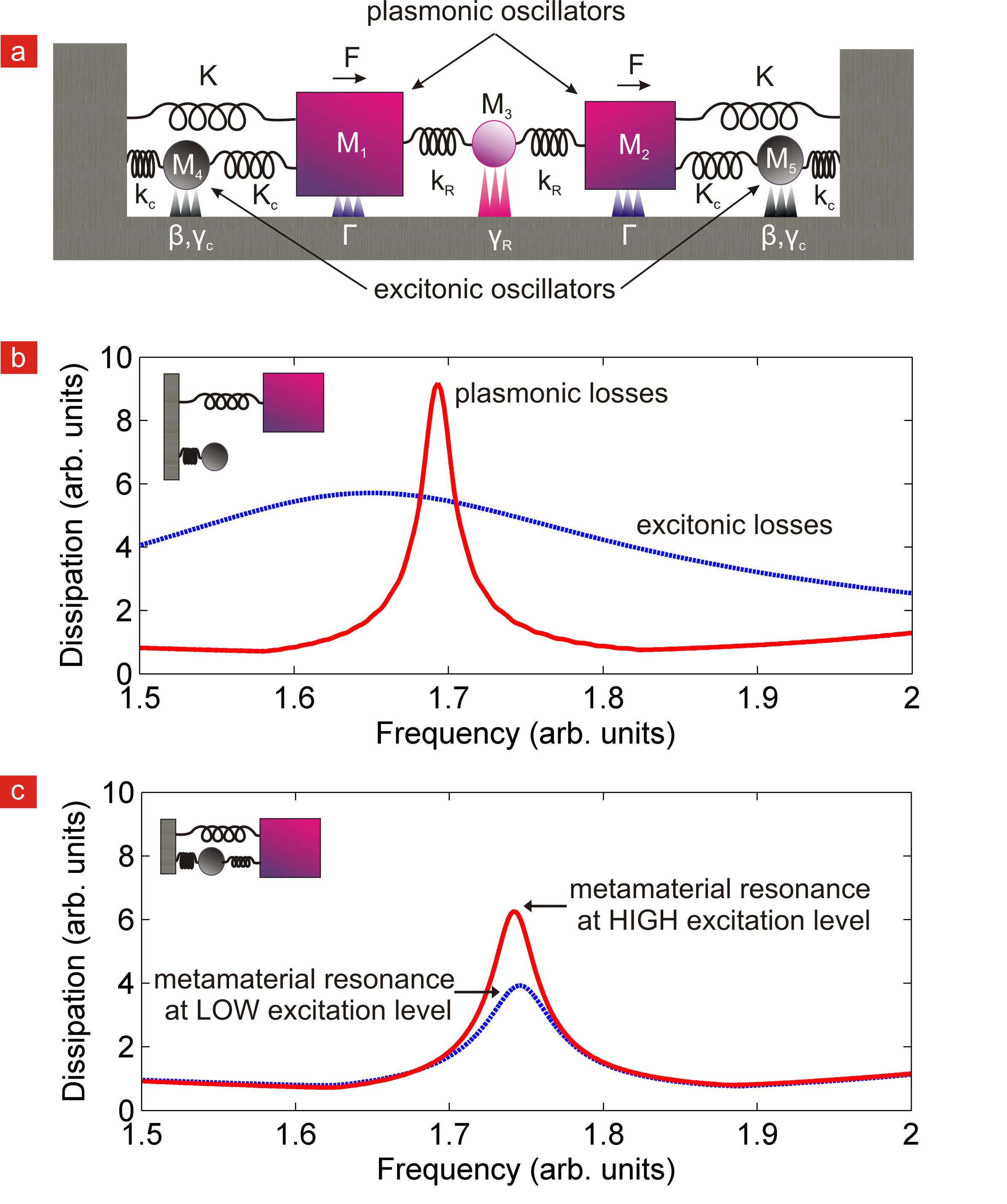}
\caption{(a) Illustrative model of the carbon nanotube metamaterial and the plasmon-exciton nonlinearity. The
metamaterial's plasmon response is represented by \emph{harmonic} oscillators with masses $M_1$ and $M_2$ linked to mass $M_3$. The
carbon nanotube's excitonic response is represented by the \emph{nonlinear} oscillators $M_4$ and $M_5$. Plate (b) separately shows the linear dissipation losses of the uncoupled plasmon (red line) and exciton (blue line) systems. Plate (c) shows the dissipation losses of the linked plasmon-exciton system at different levels of excitation. Note that the higher excitation level leads to an increase in the overall losses in spite of bleaching of the exciton absorption.}\label{model}
\end{figure}

The main features of the mechanism underlying the nonlinear optical
response of the coupled plasmon-exciton system in the carbon
nanotube metamaterial can be understood in terms of a simple
mechanical model consisting of coupled Hooke oscillators (mass on
elastic spring) driven by an external harmonic force $F$ (see Fig.
5a)~\cite{Mechanical model}. Here the plasmon
resonance is represented by the harmonic oscillation of two
masses, $M_1$ and $M_2$, elastically coupled through a third mass,
$M_3$. The oscillators represent excitations in the $\Pi$-shaped
and straight slit of the unit cell of the metamaterial and hence
have different masses, while friction $\Gamma$ stands for the
plasmonic losses. The mass $M_3$ linking the two oscillators is
also damped to account for the radiation losses $\gamma_r$. At the
plasmonic dark-mode resonance the larger masses oscillate with
\emph{opposite phases} leaving the middle mass still. Hence
radiation losses are at minimum (since the mass $M_3$ does not move), all
the energy is stored in high-amplitude oscillations of the large
masses $M_1$ and $M_2$ leading to a sharp
absorption peak associated with friction $\Gamma$ as seen in the
corresponding dissipation spectrum of Fig.~5b.
To account for the carbon nanotube layer we introduce two additional
\emph{nonlinear} oscillators containing masses $M_4$ and $M_5$. They are
responsible for the formation of the exciton absorption line (see
Fig. 5b). The saturation of the excitonic absorption in the carbon
nanotubes is introduced assuming that the oscillators are subject
to nonlinear dissipation ($\beta$). The plasmon-exciton coupling is represented by elastic springs of constant $K_c$. The model reproduces all the essential features observed in the nonlinear
response of the carbon nanotube metamaterial: When measured
separately, the plasmon resonance appears at a frequency slightly
higher then the carbon nanotube excitonic resonance and is sharper
(see Fig.~5b). For a small amplitude of the driving force $F$, corresponding to
low light intensity, the dark mode resonance experiences strong
damping as a result of the plasmon-exciton coupling. For a high level of excitation, corresponding to high levels of
light intensity, the excitonic absorption saturates and hence the
plasmonic peak, now subject to lower losses, partially recovers,
increasing in amplitude and becoming narrower (see Fig.~5b). This
illustrates our experimental observation that in the strong exciton-plasmon coupling regime (small $\delta_{pe}$) the transmission
through a metamaterial sample becomes lower at higher levels of
excitation in spite of bleaching of the exciton absorption
(compare with Fig.~4b).

In summary we have demonstrated that single walled semiconductor carbon nanotubes can be used as a very efficient agent of nonlinearity in metallic metamaterial structures. Exciton-plasmon coupling and strong resonant local fields of the metamaterial create an ultrafast nonlinear response that is at least an order of magnitude stronger than that of a bare CNT film. Importantly, the metamaterial environment allows to spectrally tailor the nonlinear response and even reverse the sign of optical nonlinearity. We argue that carbon nanotubes on metamaterials promise to offer performance that is robust, stable and free from permanent bleaching. Indeed, the resonance nonlinear properties of the suggested composite metamaterial can be easily tuned throughout the near-IR (including technologically important wavelengths such as $1.06~\mu$m and $1.55~\mu$m) by employing carbon nanotubes of different diameter and appropriately scaling the metamaterial. On the other hand, the anisotropic nature of the metamaterial offers the possibility to realize polarization sensitive nonlinearities, where nonlinear changes in transmission and reflection can even present different signs for different polarizations. This makes carbon nanotube metamaterials very promising media for various nanophotonic applications, such as optical limiting and control of laser emission.

The authors would like to acknowledge the financial support of the
Engineering and Physical Sciences Research Council (U.K.).\\

\end{document}